\documentclass[twocolumn,aps,showpacs,pre,amsmath]{revtex4}

\usepackage{amssymb}
\usepackage{graphicx}
\usepackage{dsfont}

\begin{document}

\title{Transport gap in vertical devices made of incommensurately \\ misoriented graphene layers}

\author{V. Hung Nguyen$^{1,2}$\footnote{E-mail: hung@iop.vast.ac.vn} and P. Dollfus$^1$} \address{$^1$Institut d'Electronique Fondamentale, UMR8622, CNRS, Universit$\acute{e}$ Paris Sud, 91405 Orsay, France \\ $^2$Center for Computational Physics, Institute of Physics, Vietnam Academy of Science and Technology, P.O. Box 429 Bo Ho, 10000 Hanoi, Vietnam}

\begin{abstract}
	 By means of atomistic tight-binding calculations, we investigate the transport properties of vertical devices made of two incommensurately misoriented graphene layers. With a chosen transport direction (Ox-axis), we define two classes of rotated graphene lattice distinguished by the different properties of their lattice symmetry and, hence, Brillouin zone, i.e., the two Dirac cones are located either at the same $k_y$-point ($K_y' = K_y = 0$) or at different $k_y$-points ($K_y' = -K_y = 2\pi/3L_y$, where $L_y$ is the periodic length along the Oy axis). As a consequence, a misalignment of Dirac cones of two layers occurs and a significant energy-gap ($\sim$ a few hundreds of meV) of transmission is achieved in devices made of two layers of different lattice classes. We also shown that strain engineering can be used to strongly enlarge the gap in this type of device.
\end{abstract}

\pacs{xx.xx.xx, yy.yy.yy, zz.zz.zz}
\maketitle

Graphene is nowadays one of the most attractive materials in several research fields because of its unusual, and in many respects, excellent physical properties, as a consequence of its two-dimensional honeycomb lattice. In particular, it is expected to be a good channel material in high-frequency electronic devices, flexible devices, spintronic devices, and to provide outstanding properties for photonics and optoelectronics, sensors, energy storage and conversion and so on \cite{ferr15}. However, with a view to digital applications in electronics, the lack of bandgap in graphene is known to make it difficult to turn off the current, which leads to low $I_{ON}$/$I_{OFF}$ ratio and poor current saturation \cite{meri08,alar13}, which are important limitations for innovative applications of transistors. So far, many efforts of bandgap engineering in graphene have been made to solve these issues. For instance, techniques such as cutting 2D graphene sheet into narrow nanoribbons \cite{yhan07}, depositing graphene on hexagonal boron nitride substrate \cite{khar11}, nitrogen-doped graphene \cite{lher13}, applying an electric field perpendicularly to Bernal-stacking bilayer graphene \cite{zhan09}, graphene nanomeshes \cite{jbai10}, using hybrid graphene/hexagonal boron-nitride \cite{fior12} or vertical graphene channels \cite{brit12} have been explored. Although they are certainly promising options for opening a bandgap in graphene, some of them still have their own issues while the others still need experimental verification and realization. Hence, bandgap engineering is still a timely topic for the development of graphene in nanoelectronics.

In this regard, we have recently proposed the use of strain engineering to open a finite energy gap in graphene hetero-channels \cite{hung14,chun14,hung15}. It is basically due to the fact that although it can not change the gapless character, a small strain can lead to a significant displacement of Dirac cones of the graphene's bandstructure in the k-space. Due to this effect, the misalignment of Dirac cones in different graphene sections of a heterochannel occurs and may result in a finite energy-gap of the transmission. We have hence demonstrated that strain engineering is indeed a promising option to enlarge the applications of graphene, e.g., as transistors or strain and thermal sensors.
\begin{figure}[!b]
	\centering
	\includegraphics[width=3.4in]{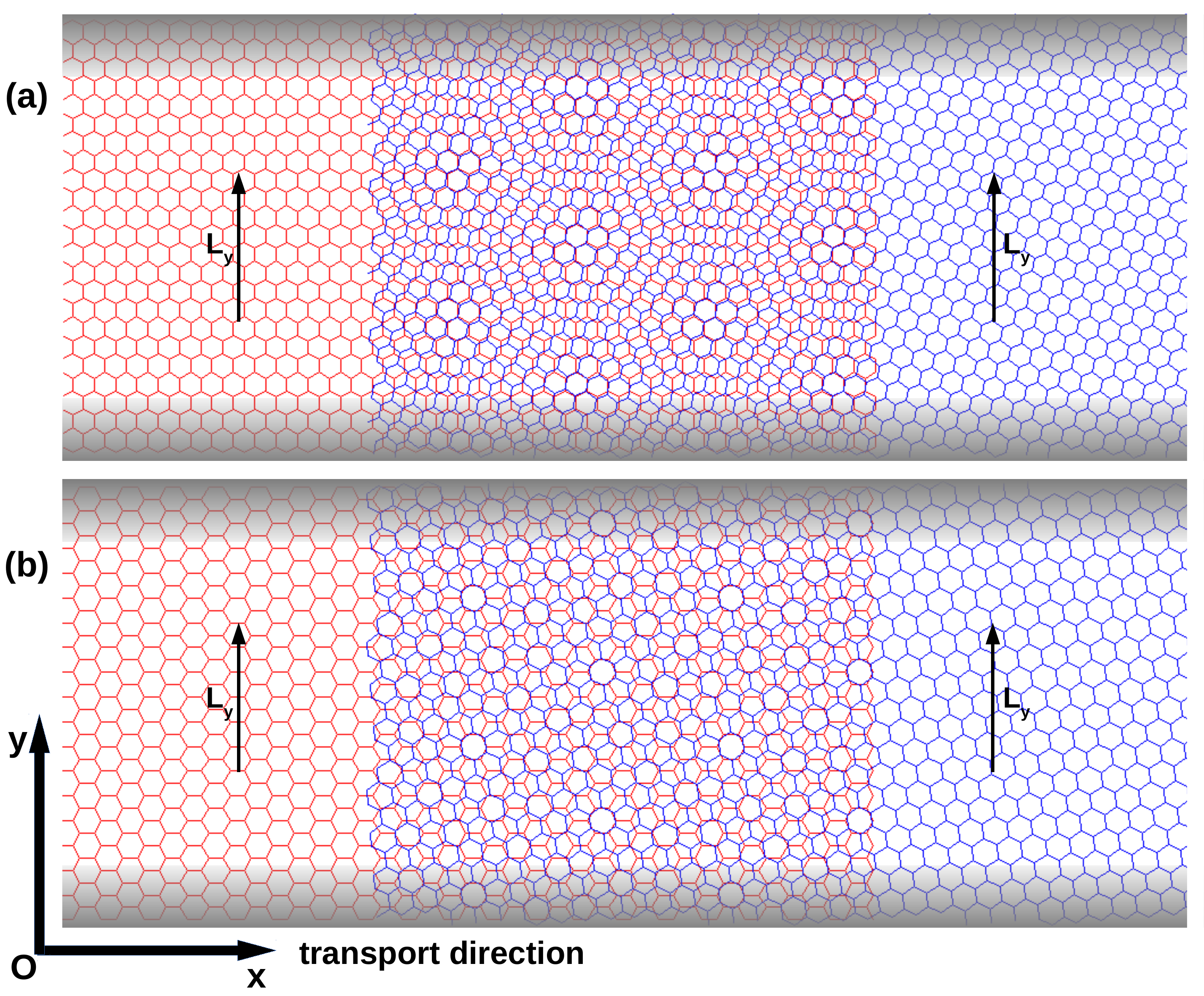}
	\caption{Schematic of devices made of vertical stack of misoriented graphene layers. The left layer has a zigzag (a) or amrchair orientation (b) along the transport direction.}
	\label{fig_sim0}
\end{figure}

\begin{figure*}[!t]
	\centering
	\includegraphics[width=5.4in]{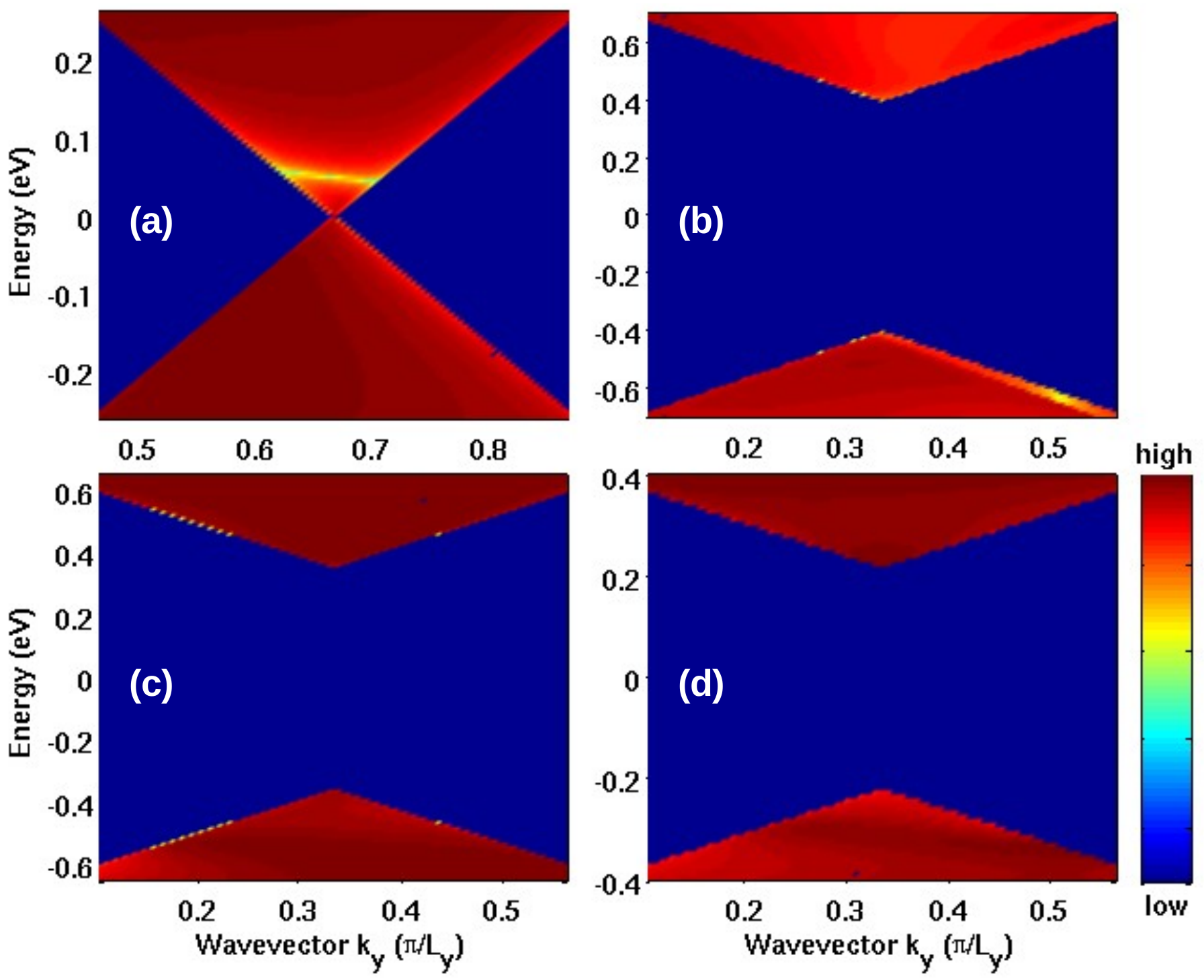}
	\caption{($E-k_y$)-maps of transmission probability around the neutrality (Dirac) point in different devices. The devices made of (AM$_{7,4}$@AM$_{4,7}$), (AM$_{6,6}$@AM$_{7,4}$), (ZZ$_{4,4}$@ZZ$_{5,3}$), (AM$_{13,7}$@AM$_{13,8}$) layers are considered in the panels (a,b,c,d), respectively (see the definition of layers in the text).}
	\label{fig_sim1}
\end{figure*}
More interestingly, devices made of vertical stack of misoriented (twisted) graphene layers have been investigated in \cite{hung15}. As suggested, this design strategy may have the advantages to be based on the use of graphene materials only and of making it possible to apply a uniform strain over the full device channel, compared to the vertical devices \cite{brit12} and strain hetero-channels \cite{hung14,chun14}. However, we have previously limited our investigation to devices made of commensurately twisted layers, an ideal case of twisted multilayer system. This naturally gives rise to a question about the transport properties of devices made of incommensurate graphene layer stacks. Hence, in the current work, we aim at answering this question and explore a new possibility of energy-gap engineering. Experimentally, controlling the misorientation of graphene layers has been explored and used to achieve resonant tunnelling in vertical devices based on graphene/hexagonal boron-nitride/graphene heterostructures \cite{mish14}. The commensurate - incommensurate transition in graphene on hexagonal boron-nitride has been also achieved in \cite{wood14}. These works could provide good suggestions for the fabrication of the structures investigated here.

\begin{figure*}[!t]
	\centering
	\includegraphics[width=5.8in]{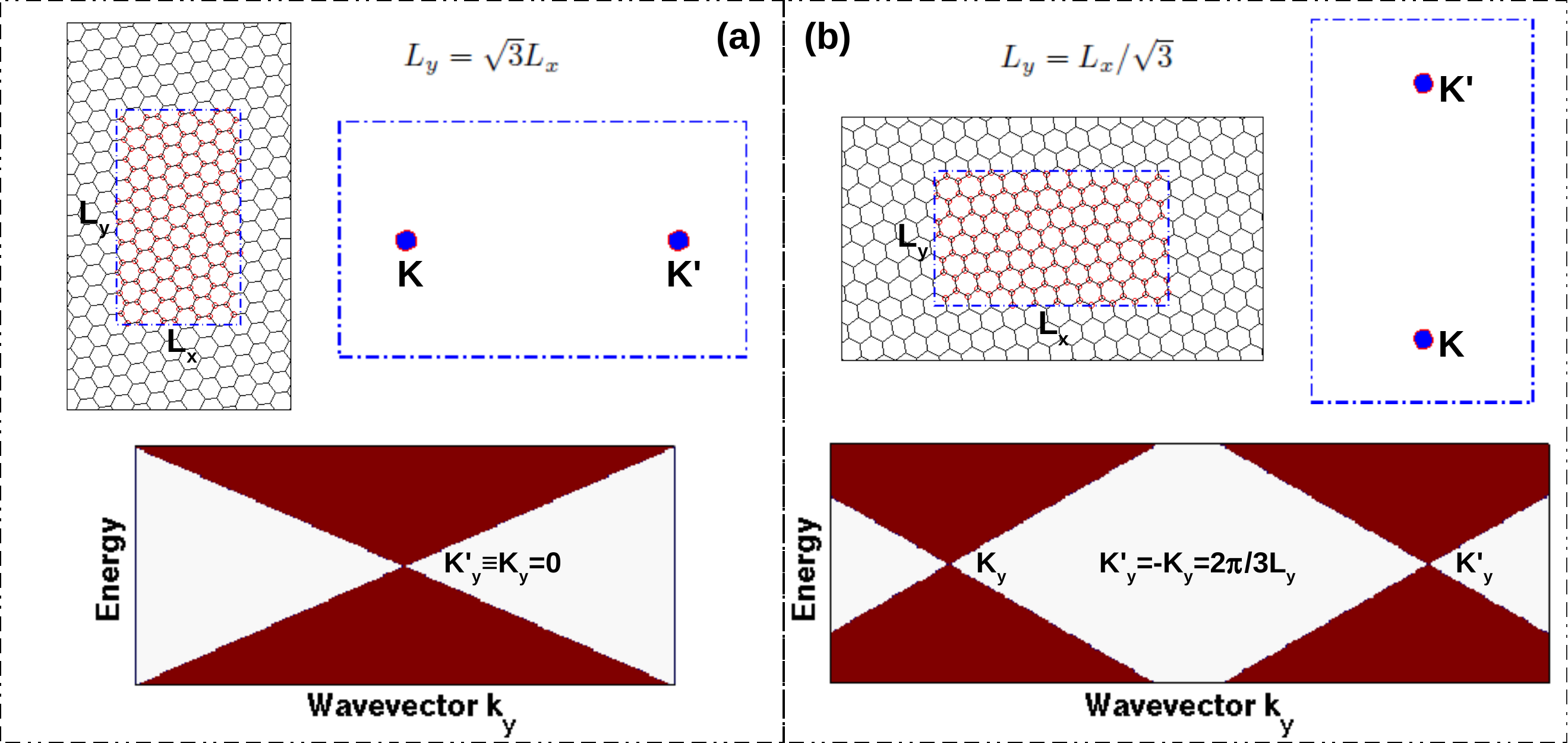}
	\caption{Schematic of two typical unit cells of graphene lattices, shape of their Brillouin zone, and corresponding pictures showing their bandstructure profile along the $k_y$ axis.}
	\label{fig_sim2}
\end{figure*}
The studied structures are schematized as in Fig. 1. They consist in two partially overlaped graphene layers and the transport (Ox) direction is perpendicular to the overlap section. The top layer is rotated relatively to the bottom one by an angle $\phi_{TL}$ but, differently from the study in \cite{hung15}, they are incommensurate layers, i.e., they do not have any common (truly) periodic lattice vector. Here, we consider 2D graphene channels, i.e., the width of graphene layers is very large, or, in other word, assumed to be infinite. Hence, to carry out the calculations, the periodicity of the channel along the transverse (Oy) direction is a necessary assumption \cite{hung14,chun14,hung15}. In this regard, we propose an approximation as follows. In all studied cases, the two layers have different lengths $L_y^1$ and $L_y^2$ of unit cells along the Oy axis. When they are coupled to form a heterochannel, we can always find two periodic vectors ${\mathbf{\tilde{L}}_y^1} = m_1 {\mathbf{L}_y^1}$ and ${\mathbf{\tilde{L}}_y^2} = m_2 {\mathbf{L}_y^2}$ that satisfy the condition $\tilde{L}_y^1 \simeq \tilde{L}_y^2$. On this basis, our calculations are performed, similarly as in \cite{hung14,chun14,hung15} with an approximately periodic length $\tilde{L}_y = ( \tilde{L}_y^1 + \tilde{L}_y^2 )/2$. This approximation can be reasonably accepted if the difference between $\tilde{L}_y^1$ and $\tilde{L}_y^2$ is small enough, as it has been confirmed in similar systems \cite{yazy10,kuma12,zhan12,fior13}. In particular, in studies on graphene grain boundaries \cite{yazy10,kuma12,zhan12}, it was shown that tight binding calculations made within the above approximation are in good agreement with density functional theory, especially in the region close to the energy-gap. The transport properties have been accurately computed around the gap even when the difference between $\tilde{L}_y^1$ and $\tilde{L}_y^2$ reaches $\sim$ 3.85 $\%$. Similar agreement has been obtained in vertical graphene/hBN heterochannels \cite{fior13}, where the lattice mismatch is about 1.8 $\%$. Throughout the present work, to ensure the accuracy, our calculations were performed with $\tilde{L}_y^1$ and $\tilde{L}_y^2$ that have only a small difference, i.e., $< 2 \%$.

Now, we would like to explain the definition of graphene lattices considered in this work. In Fig. 1, we plot the geometry of two typical devices, i.e., the left (bottom) layer has zigzag and armchair orientation in (a) and (b), respectively, with respect to the Ox axis. We distinguish two lattice types as follows. For both cases, the lattice is determined by rotating the original (zigzag or armchair) lattice by an angle $\phi_{TL}$. This angle is computed as $\cos \phi_{TL} = \frac{{\mathbf{{L_y}} \cdot \mathbf{L_y^0}}}{{\|\mathbf{L_y}\| \|\mathbf{L_y^0}\|}}$ where $\mathbf{{L_y}}$ is the vector (before the lattice is rotated) defining the size of unit cells along the Oy-axis while $\mathbf{{L_y^0}}$ is the vector of the original lattice. These vectors are determined as $\mathbf{L_y} = q_1 \mathbf{a_1} + q_2 \mathbf{a_2}$ and $\mathbf{L_y^0} = \mathbf{a_1} + \mathbf{a_2}$. For the type 1 (i.e., ZZ$_{q1,q2}$ lattice, see in Fig. 1(a)) when the original lattice has a zigzag orientation along the Ox axis, $\mathbf{a_1} = (-\sqrt{3},3)a_0/2$ and $\mathbf{a_2} = (\sqrt{3},3)a_0/2$. For the type 2 (i.e., AM$_{q1,q2}$ lattice, see in Fig. 1(b)) when the original lattice has an armchair orientation, $\mathbf{a_1} = (-3,\sqrt{3})a_0/2$ and $\mathbf{a_2} = (3,\sqrt{3})a_0/2$. Here, $a_0 = 0.142$ nm is the in-plane C-C bond length in graphene. Accordingly, we determine also the size of unit cells along the Ox axis as $\mathbf{L_x} = p_1 \mathbf{a_1} + p_2 \mathbf{a_2}$ that satisfies the condition $\mathbf{L}_x\cdot\mathbf{L}_y = 0$, i.e., $\frac{{{p_1}}}{{{p_2}}} =  - \frac{{2{q_2} + {q_1}}}{{2{q_1} + {q_2}}}$ and $\frac{{{p_1}}}{{{p_2}}} = - \frac{{2{q_2} - {q_1}}}{{2{q_1} - {q_2}}}$ for ZZ$_{q1,q2}$ and AM$_{q1,q2}$ lattices, respectively.

To compute the electronic transport in the considered systems, we employed atomistic tight-binding calculations and Green's function techniques as in \cite{hung14,chun14,hung15,pere09,moon12,wang12,kosh15,hung15b}. In particular, the hopping parameters are determined similarly as in \cite{wang12,kosh15,hung15b}, i.e., $t_{ij} = t_0 \exp \left[ -\beta \left \{r_{ij}/r_0 - 1\right\} \right]$ where $t_0 = -2.7$ \textit{eV} (resp. 0.48 \textit{eV}), $\beta = 3.37$ (resp. 7.42) and $r_0 = 1.42$ \AA{} $\equiv a_0$ (resp. 3.35 \AA{}) for in-plane (resp. interlayer) interactions.
\begin{table}[htb]
	\caption{Detailed description of some devices investigated in this work and their transport gap $E_g$:}
	\begin{center}
		\begin{tabular}{l l l l l l l}
			\hline
			\hline
			&Layer 1 \,\, &Layer 2 \,\,\,\, &$\phi_{TL}$ \,\,\,\, &\,\,$\tilde{L}_y$ ($a_0$) \,\, &$\eta$ ($\%$) \,\, &$E_g$(meV)\\
			\hline
			&\,ZZ$_{4,4}$   &\,ZZ$_{5,3}$    &8.21$^\circ$   &\,\,12.062 &1.031 &\,\,\,\,703.2 \\
			&\,ZZ$_{5,5}$   &\,ZZ$_{8,1}$    &24.18$^\circ$  &\,\,14.899 &1.351 &\,\,\,\,569.3 \\
			&\,ZZ$_{13,1}$  &\,ZZ$_{11,4}$   &11.25$^\circ$  &\,\,23.366 &0.549 &\,\,\,\,363.0 \\
			&\,ZZ$_{8,8}$   &\,ZZ$_{13,2}$   &22.94$^\circ$  &\,\,24.216 &1.790 &\,\,\,\,350.2 \\
			&\,ZZ$_{15,1}$  &\,ZZ$_{13,4}$   &9.80$^\circ$   &\,\,26.776 &0.836 &\,\,\,\,316.7 \\
			&\,ZZ$_{19,1}$  &\,ZZ$_{17,4}$   &7.78$^\circ$   &\,\,33.629 &1.061 &\,\,\,\,252.2 \\
			&\,AM$_{6,6}$   &\,AM$_{7,4}$    &25.28$^\circ$  &\,\,10.464 &1.369 &\,\,\,\,810.6 \\
			&\,AM$_{9,9}$   &\,AM$_{10,7}$   &16.99$^\circ$  &\,\,15.491 &1.250 &\,\,\,\,547.5 \\
			&\,AM$_{13,7}$  &\,AM$_{13,8}$   &5.04$^\circ$   &\,\,19.595 &0.781 &\,\,\,\,432.8 \\
			&\,AM$_{14,9}$  &\,AM$_{13,11}$  &12.41$^\circ$  &\,\,21.141 &1.342 &\,\,\,\,401.2 \\
			&\,AM$_{14,14}$   &\,AM$_{16,11}$  &17.78$^\circ$  &\,\,24.402 &1.259 &\,\,\,\,347.6 \\
			&\,AM$_{17,9}$  &\,AM$_{17,10}$  &3.87$^\circ$   &\,\,25.573 &0.458 &\,\,\,\,331.6 \\
			&\,ZZ$_{11,4}$  &\,AM$_{14,13}$  &18.59$^\circ$  &\,\,23.366 &0.549 &\,\,\,\,362.9 \\
			&\,ZZ$_{13,2}$  &\,AM$_{16,11}$  &24.83$^\circ$  &\,\,24.494 &0.500 &\,\,\,\,346.2 \\
			&\,ZZ$_{15,1}$  &\,AM$_{17,13}$  &16.20$^\circ$  &\,\,26.776 &0.836 &\,\,\,\,316.7 \\
			&\,ZZ$_{15,4}$  &\,AM$_{19,14}$  &26.22$^\circ$  &\,\,29.798 &1.689 &\,\,\,\,284.6 \\
			\hline
			\hline
		\end{tabular}
	\end{center}
	\label{tab_euler}
\end{table}

\begin{figure}[!b]
	\centering
	\includegraphics[width=3.4in]{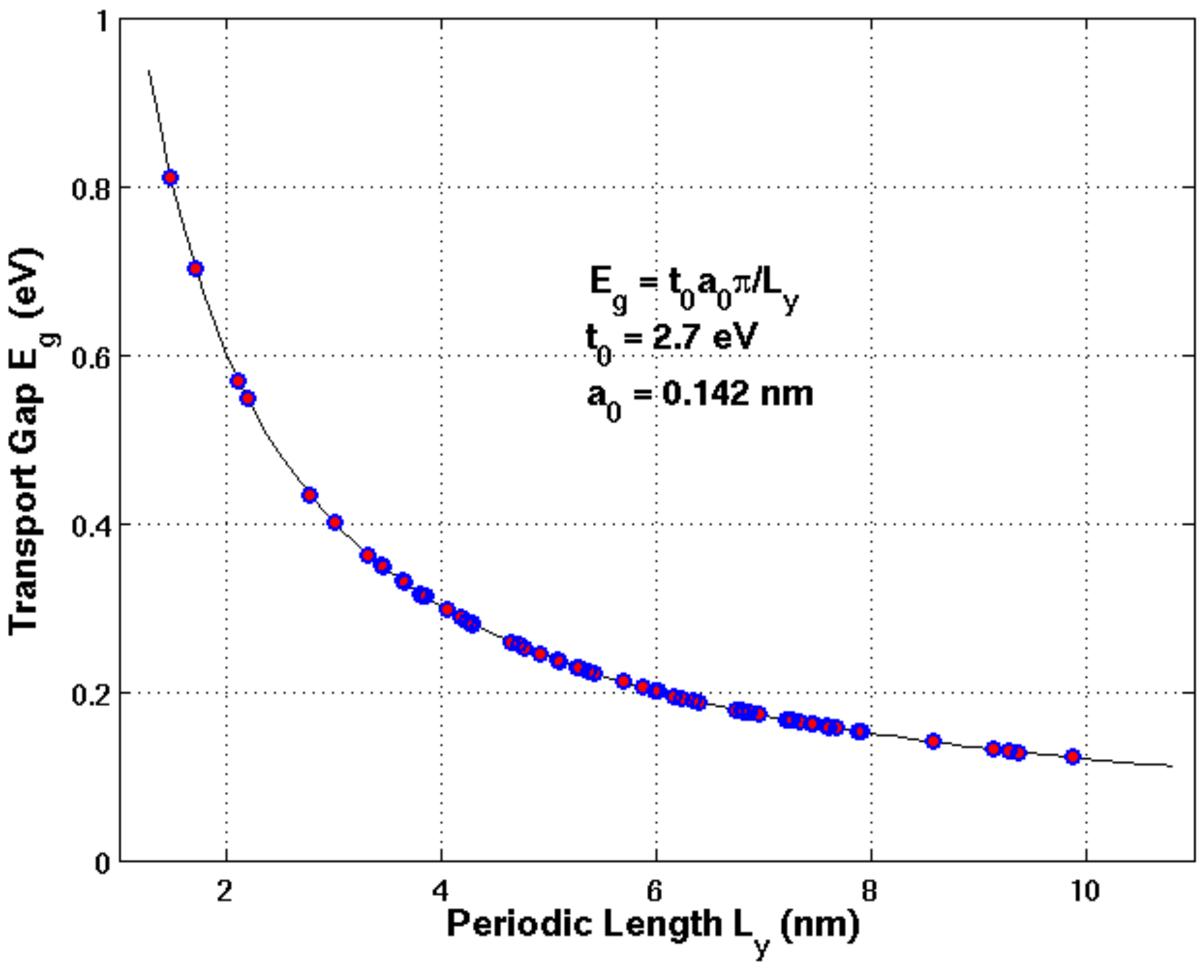}
	\caption{Transport gap as a function of periodic length $\tilde L_y$ along the Oy axis.}
	\label{fig_sim3}
\end{figure}
First, we would like to analyze some basic transport properties of the considered devices. We display in Fig. 2 the ($E-k_y$)-maps of transmision probability in four cases: the devices made of (AM$_{7,4}$@AM$_{4,7}$), (AM$_{6,6}$@AM$_{7,4}$), (ZZ$_{4,4}$@ZZ$_{5,3}$), (AM$_{13,7}$@AM$_{13,8}$) layers in the panels (a,b,c,d), respectively. Note that the two layers AM$_{7,4}$ and AM$_{4,7}$ are commensurate while the other systems are incommensurate. It is clealy shown that in the case of commensurate lattices, even if the two lattices have different orientations, there is no energy-gap of transmission. This can be explained by the fact that the commensurate lattices have the same periodic vectors and hence, because of their lattice symmetry, they have the same Brillouin zone. Since their Dirac cones are located at the same k-point, the transport gap is zero, similarly as reported in \cite{kuma12,hung15,hung15b}. The transport picture, however, can be dramatically changed in the devices made of incommensurate layers, i.e., finite energy-gaps can be achieved as shown in Figs. 2(b,c,d). The value of energy-gap is quite significant in these cases: $E_g \simeq 810.6$ meV in (b), 703.2 meV in (c), and 432.8 meV in (d). This phenomenon is explained as follows. For all cases studied, we can distinguish two classes of rotated graphene lattice as shown in Figs. 3(a) and 3(b). In the class 1, the lengths $L_x$ and $L_y$ determining the unit (smallest periodic) cell satisfy $L_y = \sqrt{3}L_x$ while $L_y = L_x/\sqrt{3}$ for the class 2. Accordingly, the Brillouin zone and Dirac cones (at the K-points) of these two lattice classes have different properties: $K_y' = K_y = 0$ for the class 1 while  $K_y' = -K_y = 2\pi/3L_y$ for the class 2 as schematized as in Fig. 3. Similarly to the case of commensurate lattices, if the channel is made of two layers of the same class, the Dirac cones of two layers are located at the same k-point and the transport gap is zero (not shown). However, if the channel is made of layers of different classes, there is a misalignment of Dirac cones of two layers in the k-space and the transport gap is finite. This essentially explains our obtained results.

In the table I, we show the detailed description of some concrete devices investigated in this work and their transport gap. $\tilde{L}_y$ is the average periodic length of two layers along the Oy axis and $\eta$ denotes the mismatch between $\tilde{L}_y^1$ and $\tilde{L}_y^2$, i.e., $\eta = |\tilde{L}_y^1 - \tilde{L}_y^2|/\tilde{L}_y$. As mentioned above, our calculations were performed with $\eta < 2 \%$ in all cases. First, we find that there are various possibilities of finite band-gap using this design strategy and $E_g$ can reach a few hundreds meV. Important, the dependence of $E_g$ on the twist angle is quite complicated but $E_g$ is inversely proportional to $\tilde{L}_y$. This is essentially explained as follows. The value of $E_g$ is basically proportional to the distance $\Delta K_y$ between Dirac cones of two layers along the $k_y$-axis \cite{kuma12,chun14,hung15}. In the considered channels, $\Delta K_y$ is actually equal to $2\pi/3\tilde{L}_y$, which fully explain the property above. On this basis, we can also analytically compute the value of $E_g$. Note that in the low energy regime (i.e., $|E| \ll 1$ eV around the Dirac point), the energy dispersion of graphene is well described by the simple formula $E(\tilde{k}) = \pm \hbar v_F \tilde{k}$ with $v_F = 3a_0t_0/2\hbar$ and $\mathbf{\tilde{k} = k - K}$ (i.e., $t_0 = 2.7$ eV and $a_0$ = 0.142 nm). The bandegde is thus defined as a function of $\tilde{k}_y$ by $E_{edge}(\tilde{k}_y) = \pm 3t_0 a_0|\tilde{k}_y|/2$. In the considered channel, the Dirac cones of two layers are located at different $k_y$-point, i.e., at $K_y^1 = 0$ and $K_y^2 = 2\pi/3\tilde{L}_y$ respectively, and the energy-gap is thus determined as the value of $E_{edge}(\tilde{k}_y)$ at the middle point \cite{kuma12,chun14,hung15}, i.e., $\tilde{k}_y = \pi/3\tilde{L}_y$ (see also in Figs. 2(b,c,d)):
\begin{equation}
   E_g = \pi\frac{a_0}{\tilde{L}_y}t_0
\end{equation}
In Fig. 4, we summarize the data obtained in all studied devices and, indeed, it confirms that the results match very well with the analytical formula above.

\begin{figure}[!t]
	\centering
	\includegraphics[width=3.4in]{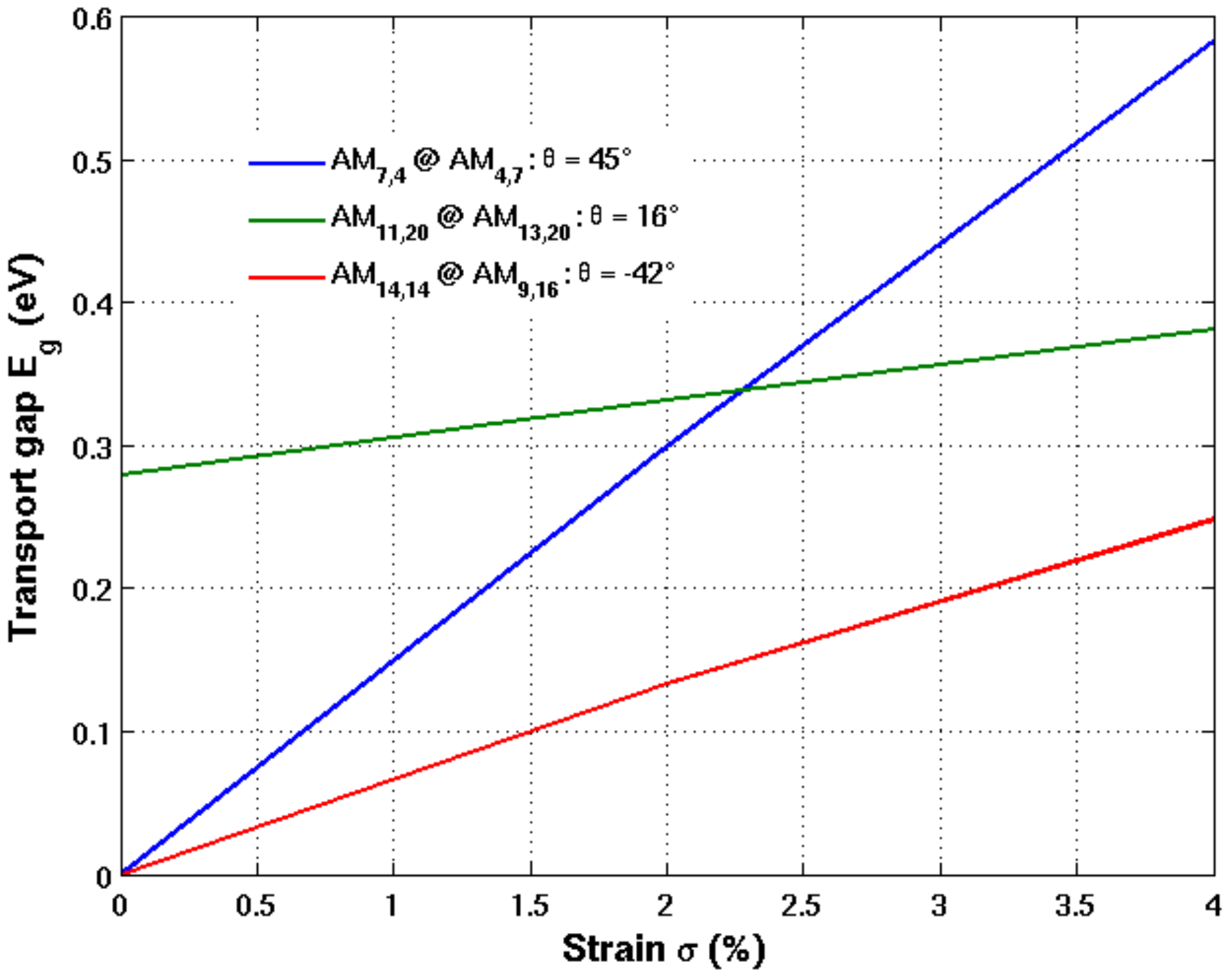}
	\caption{Strain effects on the transport gap in different devices. The strain of angle $\theta$ with respect to the transport direction (Ox) is applied \cite{hung15}.}
	\label{fig_sim3}
\end{figure}
Now, new questions arise. Since the dependence of $E_g$ on the twist angle is quite complicated, it shall be difficult to control this energy-gap in practice. Additionally, it is also difficult to control well the commensurateness of two layers. To partly solve these issues, we propose to use strain engineering as an additional ingredient. As suggested in \cite{hung15}, the strain effects are very efficient to open a finite energy-gap in devices made of commensurate lattices. Similarly, this technique can be used to open an energy-gap in the channels where two incommensurate layers are in the same lattice class, i.e., when $E_g = 0$ without strain. Additionally, in the cases of device with a small gap (i.e., if the two layers are from different lattice classes and $\tilde{L}_y$ is large), strain can be used to further enlarge $E_g$. These points are indeed shown clearly in Fig. 5 for three typical cases: commensurate (AM$_{7,4}$@AM$_{4,7}$), incommensurate (AM$_{14,14}$@AM$_{9,16}$) with zero gap and incommensurate (AM$_{11,20}$@AM$_{13,20}$) with a small gap without strain. The gap is significantly broadened, i.e., by a few hundreds meV, when a small strain of only 4 $\%$ is applied.

In conclusion, we have investigated the transport properties of vertical devices made of two incommensurately misoriented graphene layers. For a given Oxy coordinates with transport direction parallel to the Ox axis, there are two different classes of graphene lattice depending on their lattice symmetry and on the properties of their Brillouin zone. In particular, the two Dirac cones are either located at the same $k_y$-point ($K_y' = K_y = 0$) or at different $k_y$-points ($K_y' = -K_y = 2\pi/3L_y$). On this basis, a misalignment of Dirac cones of the two layers occurs and, hence, a significant energy-gap (i.e., a few hundreds of meV) of transmission is achieved in devices made of two layers of different lattice classes. The gap is shown to be inversely proportional to the periodic length $L_y$. We have also shown the possibility of using strain to enlarge the energy-gap opening in this type of device. Thus, our study suggest an alternative strategy to open an energy gap in graphene channels (without altering the graphene lattices) by controlling the misorientation of graphene layers, which should be helpful for broadening the practical applications of graphene.

\textbf{\textit{Acknowledgment.}} This research in Hanoi is funded by Vietnam's National Foundation for Science and Technology Development (NAFOSTED) under grant number 103.01-2014.24.

\end{document}